# How to Establish a Bioinformatics Postgraduate Degree Programme – A Case Study from South Africa

Philip Machanick[1] and Özlem Tastan Bishop[2,*]

Research Unit in Bioinformatics (RUBi), [1]Department of Computer Science, [2]Department of Biochemistry and Microbiology, Rhodes University, Grahamstown, 6140, South Africa

*Address correspondence to this author at Research Unit in Bioinformatics (RUBi), Department of Biochemistry and Microbiology, Rhodes University, Grahamstown, 6140, South Africa; Tel: +27-46-603-8072, Fax: +27-46-603-7576. E-mail: o.tastanbishop@ru.ac.za

<u>Philip Machanick</u> *is an associate professor of computer science at Rhodes University, South Africa. His research interests include transcription factor binding specificity, computer systems performance and computer science education.*

<u>Özlem Tastan Bishop</u> *is an associate professor and head of Research Unit in Bioinformatics (RUBi) at Rhodes University. She is president of South African Society for Bioinformatics (SASBi). Her research interests include structural bioinformatics, comparative genomics and educational approaches.*

**Submitted for publication – please check for an update that points to the published version (should it be accepted).**

# ABSTRACT

The Research Unit in Bioinformatics at Rhodes University (RUBi), South Africa offers a Masters of Science in Bioinformatics. Growing demand for Bioinformatics qualifications results in applications from across Africa. Courses aim to bridge gaps in the diverse backgrounds of students who range from biologists with no prior computing exposure to computer scientists with no biology background. The programme is evenly split between coursework and research, with diverse modules from a range of departments covering mathematics, statistics, computer science and biology, with emphasis on application to bioinformatics research. The early focus on research helps bring students up to speed with working as a researcher. We measure success of the programme by the high rate of subsequent entry to PhD study: 10 out of 14 students who completed in the years 2011-2013.

## 1. INTRODUCTION

The Research Unit in Bioinformatics at Rhodes University in South Africa (formerly Rhodes University Bioinformatics Research Group, but still abbreviated to RUBi, http://rubi.ru.ac.za) offers a one-year coursework and research thesis Masters in Bioinformatics that is (to the authors' knowledge) unique in Africa. This programme is regularly over-subscribed (e.g. 65 applications for 7 places in 2012), and has had some significant successes in producing bioinformatics researchers.

Our experiences in running this programme are that, while there are real difficulties in bridging a variety of prior backgrounds, in bringing students up to speed across a diverse range of bioinformatics skills in a short time and the lack of published curricula for a programme of this nature, the challenges are worth taking on. We use an educational approach that moves students from outsiders to practitioners in stages, recognising that integrating research into teaching is essential to rapid conversion from a variety of backgrounds with limited research experience to research-capable graduates.

The interest in this programme is driven by the growth of demand for bioinformatics research skills, the lack of similar offerings, as well as South Africa's relatively benign cost of study for students from the rest of Africa. Because of this high demand, we have been able to focus on quality, and students graduating from the programme are highly sought after for PhD study.

In the remainder of this paper, we describe our educational philosophy then describe our programme in more detail in the form of a case study. Finally, we summarise with conclusions. Since the growth of demand for bioinformatics skills is a global phenomenon, this report may be a useful guideline to other institutions wishing to establish a similar programme.

## 2. EDUCATIONAL PHILOSOPHY

A multidisciplinary programme with students from diverse backgrounds presents bigger educational challenges than a more homogeneous, focused programme. Since we are dealing with mature students who generally have at least four years of prior study, we can to some extent rely on the students' acquired study skills. However, the diverse range of origins of the students makes it difficult to rely on a consistent baseline. In South Africa, the usual route to research is to start with a three-year bachelor's degree (Bachelor of Science – BSc – in the case of science), followed by a one-year Honours degree (BSc Honours). Honours is a junior graduate degree and in most cases includes a substantial project. Prior to entry to a PhD, students usually complete a Masters degree (Master of Science, MSc) that can either be by research, with a thesis that is shorter and less demanding than for a PhD, or by a combination of coursework and research. Students from other African countries may have a different background and even within South Africa standards across universities differ widely.

Given the challenges inherent in a multidisciplinary programme compounded by the diverse institutions from which we source our students, we draw on recent advances in education theory to maximise our success rate.

Historically, much research in education has been based on cognitive models of learning. Bloom's Taxonomy [1] assumes that different levels of learning essentially amount to different degrees of cognition from factual knowledge through to critical evaluation. While we want our students to attain the higher levels of the taxonomy, knowing what these levels are does not guide us as to how to get there. More recently, the work of Piaget [2,3] has inspired *constructivism*, a view that education works best when students actively participate in applying their knowledge.

Constructivists divide learning into simpler learning that adds to the student's existing mental model, and more complex learning where the student's mental model fails to fit what is being learnt and has to be adjusted. The simpler kind of learning where new details are fitted into an existing model is often referred to as *assimilation*. Learning that requires changes to the model is referred to as *accommodation* [4].

Constructivism is also a cognitive model; it does not capture aspects of learning that are not purely cognitive, like the way practitioners in a field interact with each other, and how new knowledge becomes accepted. A newer model of learning, social construction (not to be confused with social constructivism [5], a variant on constructivism that takes into account social aspects of learning but is still a cognitive model), captures these additional aspects.

The essential idea of social construction is that the way knowledge is constructed is as important to understand as the cognitive aspects. Students may have serious misconceptions as to how knowledge is formed in an unfamiliar field [6]. Given the very diverse nature of our students, this problem is exacerbated. For this reason, we draw on the ideas of social construction theory to bring our students to a common level. Social construction is not a new idea, with roots in phenomenology [7] and the idea was known in the 1960s [8]. However, despite wide application in social sciences and areas of education as diverse as physical [9] and management education [10], social construction has not seen significant use in science education, possibly because scientists largely see their work as being in the cognitive domain.

Given the diverse nature of our students, we find it useful to think beyond the purely cognitive in our educational approach. Our goal is to produce bioinformatics researchers who know how to work with and develop knowledge, not only how to reproduce knowledge, and to be able to do so whether they are from a prior background related to biology or not. Our approach therefore has a strong component of exposing students to how the community of practice to which they aspire actually works, with frequent forays into examples of how to conduct research, how to analyze each other's work critically and how to work with research literature.

## 3. CASE STUDY: ONE-YEAR COURSEWORK AND RESEARCH THESIS BIOINFORMATICS MASTERS PROGRAMME

Although bioinformatics is a discipline on its own, it has not been established as a separate discipline at undergraduate or Honours levels at Rhodes University. It falls under Biochemistry in the Department of Biochemistry and Microbiology (BM), (previously Department of Biochemistry, Microbiology and Biotechnology, BMB). BM offers a one-year coursework and thesis bioinformatics MSc programme. This programme provides an excellent foundation for those planning to continue research in bioinformatics.

Since bioinformatics is interdisciplinary, the programme accepts students from different backgrounds, such as biochemistry, molecular biology, microbiology, zoology, physics, mathematics, statistics and computer science, and aims to bring them to an equal level of interdisciplinary knowledge. The aim is to produce MSc graduates with a strong foundation. Thus, the one-year MSc programme provides a bridging role; one end with multiple legs to get students from different disciplines, and the other end to transfer those suitable to a bioinformatics PhD degree.

## 3.1 History

At Rhodes University BMB ran a one-year coursework and research thesis MSc programme in bioinformatics in 2003 and 2004. Around that time, the South African National Bioinformatics Network (NBN) was established, and NBN set up a 6-month central bioinformatics training programme. The Rhodes University programme was therefore suspended and bioinformatics students were sent to the central programme. However NBN closed down in 2008 and the 6-months central training programme did not continue. Establishment of Rhodes University Bioinformatics Research Group (RUBi) in January 2010 and reinitiation of the one-year Bioinformatics MSc Programme in January 2011 are two important steps in the development of bioinformatics at the University. Although the MSc programme is hosted by BM, it is supported by other departments in the Science Faculty, and lecturers from Departments of Chemistry, Computer Science, Pure and Applied Mathematics, and Statistics participate in teaching and supervision (in US terminology, advising) activities. The University recognized RUBi formally as a Unit in October 2013, bringing staff members from different departments together and providing proper recognition of their contributions.

## 3.2 Overview of the programme

The Masters programme is offered over 11 months and incorporates 10 to 13 coursework modules and a research project running partially concurrently throughout the programme. The coursework modules involve an integration of formal lectures, self-learning computer-based tutorials and practicals. In addition, problem-solving tutorials are designed to guide students through current research-related problems. A number of the tutorials and practical components are assessed and contribute towards a coursework year mark. The assessments of the coursework components are through assignments, tutorials, tests, written work, practicals, and examinations. The coursework and the research project each contribute 50% to an overall mark. Successful completion of the degree is subject to a final mark of at least 50%, provided that a candidate obtains at least 50% for the coursework, with a sub-minimum of at least 40% from each module, and at least 50% for the project thesis. The coursework modules are assessed by internal grading of tutorials, assignments, tests and practicals, etc. to give a class mark, and by internal and external grading of examinations. For each module, the weighting for the class mark is 40%, and for the examination is 60%. The weightings of the various modules in the calculation of the overall coursework mark is proportional to the number of lectures given. The project is graded internally by evaluating the project proposal and presentation (10%) and project results and presentations (30%), and the thesis (60%) is graded externally by two external examiners. Ideally, at least one of the external examiners should be international.

### 3.3 Coursework component

The courses can be divided into two categories, core and supplementary (Table 1). Core courses are given every year and supplementary courses are given according to the availability and interests of lecturers. Each module is about 25 contact hours (each contact hour is 45 minutes) and presented in intensive mode over one week. The only exception is the Python programming module which was initially 4 weeks. However, our experiences showed that one full month of programming training might not be enough for some biology students. Thus, in the last two years, we introduced a basic programming module. This module is not for credit. It aims to familiarize students with core programming concepts before taking a full programming module. The first module covers Linux including bash (script) programming. The other programming modules are Matlab and R. Students are asked to apply programming skills in other modules, and are discouraged from excessive reliance on point-and-click web services. Students with no biology background attend basic biochemistry courses. Depending on their background (mathematics or computer science), their course attendance at other modules is decreased.

### 3.4 Research component

The content of each research project is determined by interests of available supervisors and Table 2, where the titles of completed theses are given, is indicative of the range of topics. Here we discuss some of the techniques used to increase research efficiency and critical thinking ability of students, which requires special care because the programme time dedicated purely to research is very limited (July – December).

*Freedom in the projects:* The backbones of the projects are given to students at the beginning of the year (February), and students are expected to build their own project proposal concurrently with taking course modules in the first part of the year. Students prepare a literature review on their research project topic and present the project proposal after passing their exams in June. The students are also encouraged to relate course modules to the project and, if relevant, to consider how and why the material applies. Thus, they have a lot of freedom in designing the details of the project. This freedom extends throughout the year while they are doing their projects as well. However the freedom has some limits, and the parameters of the project are mainly defined during one-on-one meetings with supervisors. A supervisor generally tries to steer students away from an unfeasible approach by asking questions, instead of saying "No you must not do this, do that instead". Obviously, not all MSc students are capable of designing their own research *ab initio*. To do so requires strong background knowledge and an ability for critical thinking. If they cannot build the details of the project, supervisors motivate them by suggesting various steps. On the other hand, some of them come up with very original ideas. If the students were restricted with a rigid research plan, they would not have achieved original work and would not be qualified to enter a PhD programme.

*Project presentations:* In the second part of the year, besides the project proposal presentation, students are marked for three further presentations – two progress reports and one final report. Marking criteria, with weights, are defined and applied. The marking is carried out by a combination of lecturers and supervisors, with a component of peer assessment, in which the marking is also performed by fellow students. The key advantage of peer assessment is that it helps the student markers to understand the assessment criteria for the activity, and thus, for the future, helps students to understand more precisely what is expected. Peer assessment is another example of the kind of strategy that is consistent with the social construction model of learning.

*Weekly research meetings:* It is important to spend one-on-one time with research students every week, but availability of supervisors to students is not restricted to specified times. It is very useful for supervisors to spend time in the computer laboratory, rather than sitting in their office. This encourages the students to communicate with supervisors more often. Further, since research time is very limited in a one-year MSc programme, students are asked to build the thesis as they go and send the supervisors an updated version every two weeks. Though this is a lot of work for the supervisors, we found it very useful for a number of reasons. Firstly, it keeps them focused and organized. Secondly, it helps the supervisors to see their progress and how much they grasp the methodology, analysis, progress on reviewing literature, etc.

*Journal club meetings:* In the second half of the year, at journal club meetings, every week one of the students presents a bioinformatics research article. Articles are critically discussed by the group for the methodology and research design. Initially, it is very difficult for them to pick up a good article and analyse it critically rather than simply summarizing but with time and experience they improve their critical and presentation skills to an acceptable level.

*Public bioinformatics talks and private group meetings to improve presentation skills:* Regular public talks on current bioinformatics topics or basic bioinformatics approaches help students to understand the topics and improve their presentation skills. Every bioinformatics student gives a public talk at least once during the year. After the talks, group members discuss how they found the presentation and what would improve the presentation with the main aim of improving students' presentations skills as well as learning to be a critical audience. The presenter also learns to face a critical audience and to accept criticism positively.

### 3.5  Outcomes

The main output of the programme is research-productive graduate students. The data in Table 3 indicates that over 70% (10 out of 14) of completing students have continued in PhD programmes in various South African universities (Table 3). Only one student (7%) has at time of writing moved on to bioinformatics-related employment. We have no information on the remaining 3 students. Table 2 provides the titles of the theses, indicating the content. As the research time is very limited, the projects are small and do not lead to publications immediately. When students stay on for a PhD degree, there is more opportunity to complete the project and write a paper (e.g., [11,12]).

### 3.6 Challenges

While establishing and running the programme, a number of challenges arose. Here we would like to discuss some of them with our solutions as a guide to those planning to establish a similar programme.

The major challenges in this programme can be presented under various categories.

*Required infrastructure and administrative support:* Lack of a dedicated Linux computer laboratory and adequate system administrative support presented a substantial difficulty while establishing the programme. A dedicated computer laboratory requires space and funding. Instead, a seminar room was converted into a lecture room and laptops are provided to the students. All the lectures and practicals are held in the same place. Having laptops gives excellent mobility to students. For large computational runs, server access is provided. Further, to decrease the extra load introduced to the course coordinator/faculty member due to lack of system admin support (e.g. installing the operating system to each laptop, installing and maintaining various bioinformatics related software for each module etc.), the Linux module is designed to teach students how to handle basic system administration themselves. This approach added to the skills acquired by the students.

*Range of topics:* Bioinformatics is a broad and very dynamic field. It is rapidly changing. Bioinformatics lecturers and supervisors need to follow new developments continuously. Yet, with limited staff members and time limitations, the programme is unable to cover every bioinformatics topic. Most of the specialized bioinformatics modules are designed according to the research interests of the supervisors, and further learning in these topics is supported by research projects. Although students are not exposed to a wide range of bioinformatics topics, they gain strong background in core bioinformatics courses, and they learn how to do independent research on new bioinformatics topics.

*Collaborative effort from different departments:* Teaching bioinformatics can be very daunting. Bioinformatics lecturers should have expertise in different disciplines to be able to link various aspects of the field and supervise research students. Still, beside interdisciplinary knowledge of lecturers, development of a good curriculum requires a collaborative effort from different departments [13]. This programme would not be possible without the support of various departments. This was one of the key factors while establishing the programme. The advantage of collaborative effort is that it can open opportunities to scientists to carry out interdisciplinary research, which in turn would strongly reflect back in improved teaching. The challenge was to familiarize lecturers from other research fields with bioinformatics applications. For instance, a lecturer might be excellent in teaching mathematical programming and Matlab, but he or she might not be familiar with the bioinformatics tool kit. In these cases, a bioinformatics lecturer also participates to help the Matlab lecturer bridge the two fields.

*Diversity of students:* It is challenging to take students from a diverse range of backgrounds and levels, and teach them bioinformatics in depth in a short time. Bioinformatics students need to learn a number of different subjects. If their background is biology, they need to learn basic mathematics, statistics and a computer programming language. Some students have difficulty developing their programming skills, and moving from a Windows to a Linux operating system. If their background is computer science or mathematics, they would need to learn basics of biochemistry, molecular biology and genetics. However, this diversity is used as an advantage in the programme by encouraging students with different backgrounds to work together. Furthermore, having students with different backgrounds has important impacts in the classroom, as they analyse a topic from different angles.

*Finding examiners:* Currently, one of the biggest difficulties that we have is to find external examiners to review all the research theses at once. This is important for a consistent marking process. However, each of the research theses is about 100 pages, and in practice an external examiner would be willing to review at most four theses at the same time.

## 4. CONCLUSION

There are considerable obstacles to running a successful coursework programme in bioinformatics at any level. In any rapidly-evolving multidisciplinary field, sourcing course materials and lecturers with the right skill set and the ability to relate to each other and the students is difficult. Bioinformatics is particularly challenging as comparatively few students and academics have background across all the core disciplines. At Rhodes, we have been able to overcome these difficulties by a broadly collaborative approach. Our regular journal club and research presentations help to fill gaps, and our approach to teaching that integrates research from an early stage keeps students motivated while dealing with unfamiliar material.

The social construction approach emphasises inducting students into the modes of knowledge creation in a discipline, a specially challenging problem in multidisciplinary areas where students have diverse backgrounds.

Some of our other challenges may be less universal, depending on the level of support available for infrastructure and administration. Finding examiners will be less of a problem in parts of the world that do not require external examiners for theses.

Nonetheless our key lessons in dealing with student diversity and different backgrounds of academics have wide applicability. That we are able to produce PhD-ready graduates in a year (noting that there is no coursework in a South African PhD), with our more accomplished students producing publishable research within that year or soon after, indicates that our approach is successful.

## ACKNOWLEDGEMENT

PM is funded by National Research Foundation, South Africa (grants 85362 and 78746). ÖTB thanks Sabanci University, Turkey, for hospitality. ÖTB would further like to thank to J. Baxter, R. Bernard, N. Bishop, G.L. Blatch, J. Dames, J. Gillam, G. Jäger, F. Joubert, P. Kaye, K. Lobb, S. Richner, G. Salazar and G. Wells for their initial support while establishing the programme.

**KEY POINTS:**
1. Conversion to bioinformatics from diverse backgrounds is hard: we draw on the social construction model of education to bring students up to speed. The emphasis from the start is relating knowledge to how it applies to research. In a one-year Masters programme there is only limited amount of time to convert students from diverse backgrounds to PhD-ready graduates, so we place a premium on developing research skills throughout the year, even though dedicated time for the research component only starts once the courses are completed halfway through the year.
2. Demand for bioinformatics education across Africa is high: our programme is regularly oversubscribed by a factor of 10 or more. We draw students from diverse countries across the continent, adding to the problems of diverse background.
3. Success as measured by taking up PhD study is high. Over 70% of students over the period reviewed have started a PhD. Since the programme is only a year, we take this measurement as a stronger indication of success than published papers by students.
4. Integration of a range of disciplines is a challenge. We address this by common activities such as regular journal club meetings and research presentations. Students are encouraged to use these opportunities to build their critical skills.

**Table 1:** Coursework modules

| Core modules | Content | Duration-contact hours |
|---|---|---|
| Introduction to Linux | • Linux operating system and software installation<br>• Use of Linux and Linux shell commands<br>• Application to bioinformatics problems | 20 |
| Introduction to Programming | • Basics for (Python) programming | 10 |
| Python for Bioinformatics | • Introductory and advanced Python<br>• Biopython | 75 & 1 week for assignment |
| Basic Mathematics | • Review of basic calculus<br>• Review of linear algebra | 15 |
| Mathematical Programming | • The MATLAB computational environment, MATLAB scripts, graphical output, functions, systems of linear and non-linear equations, differential equations<br>• Use of the Bioinformatics Toolbox | 20 |
| Statistics | • Introductory statistics<br>• R: statistical software | 25 |
| Basic Genomics – Part I | • Genome sequencing techniques<br>• DNA and protein databases; database searching<br>• Databases and API<br>• Sequence alignment | 25 |
| Basic Genomics – Part II | • Discovering features of interest in DNA including transcription factor binding sites<br>• Using genome browsers to obtain data<br>• Using web services and the command line to performance genome-wide and specific sequence analyses | 25 |
| Comparative Genomics | • Introduction to pairwise and multiple complete genome alignment<br>• Phylogenomics<br>• Genome evolution and horizontal gene transfer<br>• New approaches, techniques and challenges | 25 |
| Structural Bioinformatics – Part I | • Protein visualization programs;<br>• Structural biology techniques<br>• Template and non-template based protein structure prediction methods<br>• Homology modeling in detail | 25 |
| Structural Bioinformatics – Part II | • NMR<br>• Docking (Autodock)<br>• Molecular dynamics | 25 |
| **Supplementary modules** | **Content** | **Duration- contact hours** |
| Databases | • Introduction to databases<br>• Introduction to web frameworks<br>• MySQL; Django | 25 |
| Neural Network | Origins of artificial neural networks, perceptrons: their construction and deployment, convergence of perceptrons, gradient descent for optimisation, general feed-forward networks with differentiable transfer functions, backpropagation, training, assessing performance, construction and deployment of feed-forward neural networks for prediction and pattern recognition, various applications, problems. | 25 |
| Phylogenetics | Introductory phylogenetics covering neighbor-joining in detail and the principles of maximum likelihood and Bayesian inference. Bootstrap analysis, evolutionary models and comparison of tree topologies. | 25 |

**Table 2:** Research projects

| Year | Thesis title | Methodology used – Novel findings |
|---|---|---|
| 2011 | *In silico* Characterisation of the Four Canonical *Plasmodium falciparum* 70 kDa Heat Shock Proteins | Data retrieval, sequence alignment, homology modeling, protein-protein interaction analysis<br>* *A novel modeling method was suggested* |
| | Structural Analysis of Prodomain Inhibition of Cysteine Proteases in Plasmodium Species | Data retrieval, sequence alignment, phylogenetic tree analysis, homology modeling, interprotein bonding analysis<br>**Few short peptides as potential inhibitors were designed* |
| | Structural Analysis of Effects of Mutations on HIV-1 Subtype C Protease Active Site | Data retrieval, sequence alignment, homology modeling, large scale ligand docking<br>* *South African HIV-positive infants were analysed for drug related mutations* |
| 2012 | Falcipains as Malarial Drug Targets | Data retrieval, sequence alignment, phylogenetic tree analysis, homology modeling, ligand docking, protein-ligand interaction analysis<br>* *A South African natural product was identified as a potential malarial inhibitor* |
| | Sequence and Structural Investigation of the Nonribosomal Peptide Synthetases of *Bacillus atrophaeus* UCMB 5137(63Z) | Genome annotation for NRPS modules, phylogenetic studies, homology modeling and structural analysis, motif analysis<br>* *New modules were identified and linker regions were analysed* |
| | *In silico* Analysis of *Plasmodium falciparum* Hop Protein and Its Interactions with Hsp70 and Hsp90 | Data retrieval, sequence alignment, phylogenetic tree analysis, motif analysis, homology modeling, protein-protein interface analysis, alanine scanning<br>* *New Hop-Hsp90 binding region was modelled and differences between human and parasite proteins were identified* |
| | Structural Bioinformatics Analysis of Plasmodium DNAJ Proteins and the Interactions with *Plasmodium falciparum* Hsp70s | Data retrieval, large scale sequence alignment, grouping, homology modeling, docking, protein-protein interaction analysis<br>* *New Plasmodium DNAJ proteins identified* |
| | Structural Bioinformatics Analysis of the HSP40 and HSP70 Molecular Chaperones from Humans | Data retrieval, large scale sequence alignment, grouping, homology modeling, docking, protein-protein interaction analysis<br>* *Detailed analysis between Hsp40 and Hsp70 was done* |
| | Influence of Non-Synonymous Sequence Mutations on the Architecture of HIV-1 Clade C Protease Receptor Site: Docking and Molecular Dynamics Studies | Data retrieval, sequence alignment, homology modeling, large scale ligand docking, protein-ligand interaction analysis, preliminary MD calculations<br>* *Analysis on differences on drug resistance between Clade B and Clade C showed interesting results* |
| | A central enrichment-based comparison of two alternative methods of generating transcription factor binding motifs from protein binding microarray data | Data retrieval, motif enrichment analysis<br>* *Differences in motif quality across two competing databases measured* |
| 2013 | A Step Forward in Defining Hsp90s as Potential Drug Targets for Human Parasitic Diseases | Data retrieval, analysis of physicochemical properties (in large scale), phylogenetic tree calculations, motif analysis (script based analysis), SCA analysis (co-evolution studies)<br>* *Some important differences between the parasite and human Hsp90 proteins were identified* |
| | Large Scale Bioinformatics Analysis of Auxiliary Activity Family 9 Enzymes | Data retrieval, sequence alignment, phylogenetic tree analysis, motif analysis, homology modeling<br>* *A novel sub-type group is identified* |
| | Transcription factor binding: Investigating the role of distance between transcription factor binding site and transcription start site | Data retrieved, statistical analysis<br>* *Differential locality of transcription finding with respect to transcription start site measured* |
| | Analysis of transcription factor binding specificity using ChiP-seq data | Data retrieval, differential motif enrichment analysis<br>* *Differential motif enrichment across cell lines measured* |

**Table 3:** Details of the MSc students

| Year | Student number | Country | Gender | Background | Financial support | PhD continuation | Bioinformatics related employment |
|---|---|---|---|---|---|---|---|
| **2011** | 3 | 1 South African<br>2 Kenyan | 1 Female<br>2 Male | 1 Biochemistry<br>1 Biomedical Technology<br>1 Pharmacy | All University or government supported | 1 at Rhodes University | 1 in Kenya |
| **2012** | 7 | 3 South African<br>3 Kenyan<br>1 Nigerian | 4 Female<br>3 Male | 3 Biochemistry<br>1 Biochemistry/Microbiology<br>1 Medical Laboratory Science<br>1 Medical Microbiology<br>1 Zoology | All University or government supported | 2 at Rhodes University<br>1 at University of Pretoria<br>2 at University of Western Cape<br>1 at University of the Witwatersrand | |
| **2013** | 4 | 2 South African<br>1 Kenyan<br>1 Zimbabwean | 4 Male | 2 Biochemistry<br>1 Biotechnology<br>1 Mathematics/Comp Science | All University or government supported | 3 at Rhodes University | |
| **2014** | 8 | 4 South African<br>1 from Botswana<br>1 Lesoto<br>2 Zimbabwean | 2 Female<br>6 Male | 1 Biochemistry<br>2 Biochemistry/Microbiology<br>1 Biotechnology<br>1 Chemistry<br>1 Mathematics/Comp Science<br>1 Microbiology<br>1 Molecular and Cell Biology | 6 University or government supported<br>2 self supported | N/A | N/A |